\documentclass[10pt, conference, compsocconf]{IEEEtran}
\usepackage{amssymb}
\usepackage{url}
\usepackage{graphicx}
\usepackage{subfigure}
\usepackage{epsfig}
\usepackage{times}
\usepackage{color}
\usepackage{multirow}
\usepackage{xspace}
\usepackage{algorithm}
\usepackage{algorithmic}
\usepackage{afterpage}

\graphicspath{{FIG/}}
\begin{document}

\newcommand{\hide}[1]{}
\newcommand{\semihide}[1]{{\tiny #1}}
\newcommand{\reminder}[1]{\textsf{{\textcolor{red}{[[#1]]}}}}
\newcommand{\rev}[1]{\textbf{*** #1}}
\newcommand{\xhdr}[1]{\vspace{1mm}\noindent{{\bf #1.}}}
\newcommand{\denselist}{ \itemsep -1pt\topsep -1pt\partopsep -1pt }
\newcommand{\ie}{\emph{i.e.}}
\newcommand{\eg}{\emph{e.g.}}
\newcommand{\NodeSwap}{\textsc{NodeSwap}\xspace}
\newcommand{\Random}{\textsc{Random}\xspace}
\newcommand{\Expand}{\textsc{Expand}\xspace}
\newcommand{\Shrink}{\textsc{Shrink}\xspace}
\newcommand\blfootnote[1]{%
  \begingroup
  \renewcommand\thefootnote{}\footnote{#1}%
  \addtocounter{footnote}{-1}%
  \endgroup
}

\title{Defining and Evaluating Network Communities
based on Ground-truth}
\author{\IEEEauthorblockN{Jaewon Yang}
\IEEEauthorblockA{
Stanford University\\
crucis@stanford.edu}
\and
\IEEEauthorblockN{Jure Leskovec}
\IEEEauthorblockA{
Stanford University\\
jure@cs.stanford.edu}
}
\maketitle

\begin{abstract}
Nodes in real-world networks 
organize into densely linked communities where edges appear with high concentration among the members of the community. Identifying such communities of nodes has proven to be a challenging task mainly due to a plethora of definitions of a community, intractability of algorithms, issues with evaluation and the lack of a reliable gold-standard ground-truth.

In this paper we study a set of 230 large real-world social, collaboration and information networks where nodes explicitly state their group memberships. For example, in social networks nodes explicitly join various interest based social groups. We use such groups to define a reliable and robust notion of {\em ground-truth communities}.

We then propose a methodology which allows us to compare and quantitatively evaluate how different structural definitions of network communities correspond to ground-truth communities. We choose 13 commonly used structural definitions of network communities and examine their sensitivity, robustness and performance in identifying the ground-truth.
%
%
We show that the 13 structural definitions are heavily correlated and naturally group into four classes. We find that two of these definitions, Conductance and Triad-participation-ratio, consistently give the best performance in identifying ground-truth communities.
%
%
%
%
We also investigate a task of detecting communities given a single seed node.  We extend the local spectral clustering algorithm into a heuristic parameter-free community detection method that easily scales to networks with more than hundred million nodes. The proposed method achieves 30\% relative improvement over current local clustering methods.



\end{abstract}

\let\thefootnote\relax\footnotetext{This paper has been published in the Proceedings of 2012 IEEE International Conference on Data Mining (ICDM), 2012.} 


\section{Introduction}
\label{sec:intro}
Networks are a natural way to represent social~\cite{mislove07measurement}, biological~\cite{palla05_OveralpNature}, technological~\cite{jure07viral}, and information~\cite{flake00_efficient} systems. Nodes in such networks organize into densely linked groups that are commonly referred to as {\em network communities}, clusters or modules~\cite{newman02community}. There are many reasons why nodes in networks organize into densely linked clusters. For example, society is organized into social groups, families, villages and associations~\cite{feld86focused,granovetter73ties}. On the World Wide Web, topically related pages link more densely among themselves~\cite{flake00_efficient}. And, in metabolic networks, densely linked clusters of nodes are related to functional units, such as pathways and cycles \cite{palla05_OveralpNature}.

%

To extract communities from a given undirected network, one typically chooses a scoring function (\eg, modularity) that quantifies the intuition that communities correspond to densely linked sets of nodes. Then one applies a procedure to find sets of nodes with a high value of the scoring function. Identifying such communities in networks~\cite{karypis98_metis,dhillon07graclus,Schaeffer07_survey,fortunato09community} has proven to be a challenging task~\cite{Fortunato07_ResolutionPNAS,jure08ncp2,jure10community} due to three reasons: There exist multiple structural definitions of network communities~\cite{danon05community,RCCLP04_PNAS}; Even if we would agree on a single common structural definition (\ie, a single scoring function), the formalizations of community detection lead to NP-hard problems~\cite{Schaeffer07_survey}; And, the lack of reliable ground-truth makes evaluation extremely difficult.


Currently the performance of community detection methods is evaluated by manual inspection. For each detected community an effort is made to interpret it as a ``real'' community by identifying a common property or external attribute shared by all the members of the community. For example, when examining communities in a scientific collaboration network, we might by manual inspection discover that many of detected communities correspond to groups of scientists working in common areas of science~\cite{newman04community}.
Such anecdotal evaluation procedures require extensive manual effort, are non-comprehensive and limited to small networks.

%

A possible solution would be to find a reliable definition of explicitly labeled gold-standard ground-truth communities. Using such ground-truth communities would allow for {\em quantitative} and {\em large-scale} evaluation and comparison of network community detection methods. Such ability would enable the field to move beyond the current standard of anecdotal evaluation of communities to a comprehensive evaluation of community detection methods based on their performance to extract the ground-truth.

The contributions of our work are three fold. First, we describe a set of 230 large social and information networks where we define ground-truth communities in a reliable way. Second, based on the ground-truth we quantitatively evaluate 13 commonly used structural definitions of network communities and examine their robustness and sensitivity to noise. Third, we extend the local spectral clustering algorithm into a parameter-free community detection method that scales to networks of hundreds of millions of nodes.




\xhdr{Present work: Ground-truth communities}
Next we describe the proposed definition of ground-truth communities and argue why it corresponds to ``real'' communities.

Generally, after communities are identified based on the structure of given network, the essential next step is to interpret them by identifying a common external property or function that the members of a given community share and around which the community organizes~\cite{feld86focused}. For example, given a protein-protein interaction network of a cell we identify communities based on the structure of the network and then find that these communities correspond to real functional units of a cell. Thus, the goal of community detection is to identify sets of nodes with a common (often external/latent) function based only the connectivity structure of the network. A {\em common function} can be common role, affiliation, or attribute~\cite{granovetter73ties}. In our protein interaction network example above, such common function of nodes would be `belonging to the same functional unit'. Community detection methods identify communities based on {\em structure} while the extracted communities are evaluated based on their {\em function}. So we distinguish between structural and functional definitions of communities. We use common function of nodes to define ground-truth communities.





\xhdr{Present work: Networks with ground-truth}
We gathered 230 networks from a number of different domains and research areas where nodes explicitly state their ground-truth community memberships.
Our collection consists of social, collaboration and information networks for each of which we find a robust functional definition of ground-truth.

For example, in online social networks (like, Orkut, LiveJournal, Friendster and 225 different Ning networks) we consider explicitly defined {\em interest based groups} (\eg, fans of Lady Gaga, students of the same school) as ground-truth communities. Nodes explicitly join such groups that organize around specific topics, interests, and affiliations~\cite{feld86focused,granovetter73ties}.
Next, we also consider the Amazon product co-purchasing network where we define ground-truth using hierarchically nested product categories. Here all members (\ie, products) of the same ground-truth community share a common function or purpose.
Last, in the scientific collaboration network of DBLP we use publication venues as proxies for ground-truth research communities. Our reasoning here is that in scientific collaboration networks, real communities would correspond to areas of science. Thus, we use journals and conferences as proxies for scientific communities.



%
%

\xhdr{Present work: Methodology and findings}
%
The ground-truth allows us to examine how well various structural definitions of network communities correspond to real functional groups (\ie, ground-truth communities). A good structural definition of a community would be such that it would detect connectivity patterns that correspond to real groups (\ie, the ground-truth). This means that we can evaluate different structural definitions based on their ability to identify connectivity structure of ground-truth communities.

We study 13 commonly used structural definitions of communities and examine their quality, sensitivity and robustness. Each such definition corresponds to a scoring function that scores a set of nodes based on their connectivity. A high score means that a set of nodes closely resembles the connectivity communities.
By comparing correlations of scores that different structural definitions assign to ground-truth communities, we find that the 13 definitions naturally group into four distinct classes These classes correspond to definitions that consider: (1) only internal community connectivity, (2) only external connectivity of the nodes to the rest of the network; (3) both internal and external community connectivity, and (4) network modularity.

We then consider an axiomatic approach and define four intuitive properties that communities would ideally have. Intuitively, a ``good'' community is cohesive, compact, and internally well connected while being also well separated from the rest of the network. This allows us to characterize which connectivity patterns a given structural definition detects and which ones it misses. We also investigate the robustness of community scoring functions based on four types of randomized perturbation strategies.
Overall, evaluation shows that scoring functions that are based on triadic closure~\cite{watts98collective} and the conductance score~\cite{ShiMalik00_NCut} best capture the structure of ground-truth communities. 
%


Last, we also investigate a task of detecting communities from a single seed node. The task is to discover all members of a community from a single seed member node. We extend the local spectral clustering algorithm~\cite{andersen06local} into a parameter-free community detection method that scales to networks of hundreds of millions of nodes. Our method recovers ground-truth communities with 30\% relative improvement in the F1-score over the current local graph partitioning methods.



To the best of our knowledge our work is the first to use social and information networks with explicit community memberships to define an evaluation methodology for comparing network community detection methods based on their accuracy on real data. 
We believe that the present work will bring more rigor to the standard for the evaluation of community detection methods. All our datasets can be downloaded at \url{http://snap.stanford.edu}.



\section{Community scoring functions\\ and data sets}
\label{sec:data}

We start by describing the network datasets and our proposed functional definitions of ground-truth communities. Then we continue with outlining 13 commonly used structural definitions of network communities.

\begin{table}[t]
\centering
\footnotesize
    \begin{tabular}{l||r|r|r|r|r}
    Dataset & $N$ & $E$ & $C$ & $S$ & $A$ \\ \hline \hline
    LiveJournal&4.0M&34.9M&311,782&40.06&3.09\\ \hline
    Friendster&117.7M&2,586.1M&1,449,666&26.72&0.32\\ \hline
    Orkut&3.0M&117.2M&8,455,253&34.86&95.9\\ \hline
    Ning (225 nets)&7.0M&35.5M&137,177&46.89&0.92\\ \hline
    Amazon&0.33M&0.92M&49,732&99.86&14.83\\ \hline
    DBLP&0.42M&1.34M& 2,547&429.79&2.56
    \end{tabular}
    \vspace{-2mm}
    \caption{230 social, collaboration and information networks with explicit ground-truth communities. $N$: number of nodes, $E$: number of edges, $C$: number of communities, $S$: average community size, $A$: community memberships per node. Ning statistics are aggregated over 225 different subnetworks.}
    \label{tab:data}
    \vspace{-10mm}
\end{table}

\xhdr{Networks with ground-truth communities}
%
%
%
Overall we consider 230 large social, collaboration and information networks, where for each network we have a graph and a set of functionally defined ground-truth communities.
Members of these ground-truth communities share a common function, property or purpose. Networks that we study come from a wide range of domains and sizes. 
Table~\ref{tab:data} gives the networks.

First, we consider online social networks (the LiveJournal blogging community \cite{lars06groups}, the Friendster online network \cite{mislove07measurement}, and the Orkut social network \cite{mislove07measurement}) where users create explicit functional groups to which others then join and share content. These groups are created based on specific topics, interests, hobbies and geographical regions. For example, LiveJournal categorizes groups into the following types: culture, entertainment, expression, fandom, gaming, life/style, life/support, sports, student life and technology. Similarly, in other social networks considered in this study users define topical communities that others then join. 
We consider each such explicit interest-based group as a ground-truth community.
%
%
Similarly, we have a set of 225 different online social networks \cite{kairam12ning} that are all hosted by the Ning platform. It is important to note that each Ning network is a separate social network --- an independent website with a separate user community. For example, the NBA team Dallas Mavericks
and diabetes patients network TuDiabetes all use Ning to host their separate online social networks. After joining a specific network, users then create and join groups. For example, in TuDiabetes, Ning network groups form around specific types of diabetes, different age groups, and similar.
Note that these are exactly the properties around which we expect communities to form in a network of diabetes patients. Again, we use such explicitly defined functional groups as ground-truth communities.


The second type of network we consider is the Amazon product co-purchasing network~\cite{jure07viral}. The nodes of the network represent products and edges link commonly co-purchased products. Each product (\ie, node) belongs to one or more hierarchically organized product categories and products from the same category define a group which we view as a ground-truth community. Note that here the definition of ground-truth is somewhat different. In this case, nodes that belong to a common ground-truth community share a common function or purpose.


Finally, we also consider the DBLP scientific collaboration network~\cite{lars06groups} where nodes represent authors and edges connect authors that have co-authored a paper. To define ground-truth in this setting we reason as follows. Communities in a scientific domain correspond to people working in common areas and subareas of science. However, note that publication venues serve as good proxies for scientific areas: People publishing in the same conference form a scientific community. Thus we use publication venues (\ie, conferences) as ground-truth communities which serve as proxies for highly overlapping scientific communities around which the collaboration network then organizes.

All our networks and the corresponding ground-truths are complete and publicly available at \url{http://snap.stanford.edu/data}.
The results we present here are consistent and robust across a wide range of networks and across an even wider range of groups. This gives further evidence that our approach is general and well-founded. Our work is consistent with the premise that is implicit in all community detection works: members of structural communities share some functional role or property that serves as an organizing principle of the network. Here we use functionally defined groups as labeled ground-truth communities.

Note that our work is fundamentally different from Ahn et al.~\cite{Ahn10LinkCommunitiesNature}, who evaluated communities with attribute based node-node similarity of the members. This approach, for example, folds all social dimensions (family, school, interests) around which separate communities form into one similarity metric~\cite{mcpherson83blau}. In contrast, we do not use node similarity to define communities. Rather, we harness explicitly labeled functional groups as labels of ground-truth communities.







\xhdr{Data preprocessing}
%
To represent all networks in a consistent way we consider each network as an unweighted undirected static graph. Because members of the group may be disconnected in the network, we consider each connected component of the group as a separate ground-truth community. However, we allow ground-truth communities to be nested and to overlap.


\xhdr{Community scoring functions}
We now proceed to discuss various scoring functions that characterize how community-like is the connectivity structure of a given set of nodes. The idea is that given a community scoring function, one can then find sets of nodes with high score and consider these sets as communities. All scoring functions build on the intuition that communities are sets of nodes with many connections between the members and few connections from the members to the rest of the network. There are many possible ways to mathematically formalize this intuition. We gather 13 commonly used scoring functions, or equivalently, 13 structural definitions of network communities. Some scoring functions are well known in the literature, while others are proposed here for the first time.

Given a set of nodes $S$, we consider a function $f(S)$ that characterizes how community-like is the connectivity of nodes in $S$. Let $G(V,E)$ be an undirected graph with $n = |V |$ nodes and $m = |E|$ edges. Let $S$ be the set of nodes, where $n_S$ is the number of nodes in $S$, $n_S = |S|$; $m_S$ the number of edges in $S$, $m_S = |\{(u, v) \in E : u \in S, v \in S\}|$; and $c_S$, the number of edges on the boundary of $S$, $c_S = |\{(u, v) \in E: u \in S, v \not \in S\}|$; and $d(u)$ is the degree of node $u$. We consider 13 scoring functions $f(S)$ that capture the notion of quality of a network community $S$. The experiments we will present later reveal that scoring functions naturally group into the following four classes:

{\bf \em (A) Scoring functions based on internal connectivity:}
\begin{itemize}
  \denselist
  \item \textbf{Internal density:} $f(S) = \frac{m_S}{n_S(n_S-1)/2}$ is the
      internal edge density of the node set $S$~\cite{RCCLP04_PNAS}.
  \item \textbf{Edges inside:} $f(S) = m_S$ is the number of edges between the
      members of $S$~\cite{RCCLP04_PNAS}.
  \item \textbf{Average degree:} $f(S) = \frac{2 m_S}{n_S}$ is the average internal degree
        of the members of $S$~\cite{RCCLP04_PNAS}.
  \item \textbf{Fraction over median degree (FOMD):}\\
  $f(S) = \frac{|\{u:u \in S, |\{(u,v):v \in S\}| > d_m\}|}{n_S}$
        is the fraction of nodes of $S$ that have internal degree higher than
        $d_m$, where $d_m$ is the median value of $d(u)$ in $V$.
  \item \textbf{Triangle Participation Ratio (TPR):}\\
  $f(S) = \frac{|\{u:u \in S, \{(v,w):v,w \in S, (u,v) \in E, (u,w) \in E, (v,w) \in E\} \neq \varnothing\}|}{n_S}$ is the fraction of nodes in $S$ that belong to a triad.
\end{itemize}

{\bf \em (B) Scoring functions based on external connectivity:}
\begin{itemize}
  \denselist
  \item \textbf{Expansion} measures the number of edges
      per node that point outside the cluster: $f(S) = \frac{c_S}{n_S}$~\cite{RCCLP04_PNAS}.
  \item \textbf{Cut Ratio} is the fraction of existing edges (out of all possible edges) leaving the cluster: $f(S) = \frac{c_S}{n_S(n-n_S)}$~\cite{fortunato09community}.
\end{itemize}

{\bf \em (C) Scoring functions that combine internal and external connectivity:}
\begin{itemize}
  \denselist
  \item \textbf{Conductance:} $f(S) = \frac{c_S}{2m_S+c_S}$ measures the fraction of
      total edge volume that points outside the cluster~\cite{ShiMalik00_NCut}.
  \item \textbf{Normalized Cut:} $f(S) = \frac{c_S}{2m_S+c_S} + \frac{c_S}{2(m-m_S)+c_S}$~\cite{ShiMalik00_NCut}.
  \item \textbf{Maximum-ODF (Out Degree Fraction):} \\$f(S) = \max_{u \in S}
      \frac{|\{(u,v) \in E:v \not\in S\}|}{d(u)}$ is the maximum fraction of
      edges of a node in $S$ that point outside $S$~\cite{flake00_efficient}.
  \item \textbf{Average-ODF:} $f(S) = \frac{1}{n_S} \sum_{u \in S}\frac{|\{(u,v) \in E:v
      \not\in S\}|}{d(u)}$) is the average fraction of edges of nodes in $S$ that
      point out of $S$~\cite{flake00_efficient}.
  \item \textbf{Flake-ODF:} $f(S) = \frac{|\{u:u \in S,|\{(u,v) \in E:v \in
      S\}|<d(u)/2\}|}{n_S}$ is the fraction of nodes in S that have fewer
      edges pointing inside than to the outside of the cluster~\cite{flake00_efficient}.
\end{itemize}

{\bf \em (D) Scoring function based on a network model:}
\begin{itemize}
  \denselist
  \item \textbf{Modularity:} $f(S) = \frac{1}{4} (m_S - E(m_S))$ is the
      difference between $m_S$, the number of edges between nodes in $S$ and $E(m_S)$, the
      expected number of such edges in a random graph with identical degree sequence~\cite{newman2006_ModularityPNAS}.
\end{itemize}

\begin{figure}[t]
	\centering
    \includegraphics[width=0.3\textwidth]{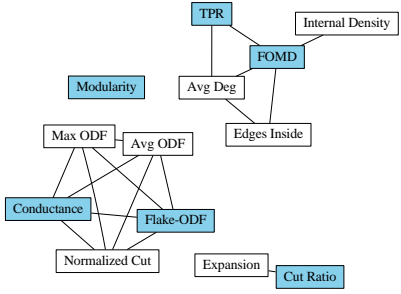}
    \vspace{-3mm}
	\caption{Clusters based on correlations of community scoring functions.}
    \label{fig:graph_scores}
    \vspace{-5mm}
\end{figure}

\xhdr{Experimental result: Four classes of scoring functions}
Next we examine relationship the 13 community scoring functions we introduced. For each of the 10 million ground-truth communities in our networks, we compute a score using each of the 13 scoring functions. We then create a correlation matrix of scoring functions and threshold it. Fig.~\ref{fig:graph_scores} shows connections between scoring functions with correlation $\ge 0.6$ (on the LiveJournal network).
We observe that scores naturally group into four clusters. This means that scoring functions of the same cluster return heavily correlated values and quantify the same aspect of connectivity structure. Overall, none of the scoring functions are negatively correlated, which means that none of them systematically disagree. Interestingly, Modularity is not correlated with any other scoring function (Avg. degree is the most correlated at 0.05 correlation). We observe similar results in other all data sets.

The experiment demonstrates that even though many different structural definitions of communities have been proposed, these definitions are heavily correlated. Essentially there are only 4 different structural notions of network communities as revealed by Fig.~\ref{fig:graph_scores}. For brevity in the rest of the paper we present results for 6 representative scoring functions (denoted as blue nodes in Fig.~\ref{fig:graph_scores}): 4 from the two large clusters and 2 from the two small clusters).

We also note that here we computed the values of the 13 scores on ground-truth communities. In reality the aim of community detection is to find sets of nodes that maximize a given scoring function. Exact maximization of these functions is typically NP-hard and leads to its own set of interesting problems. (Refer to~\cite{jure10community} for discussion.)



\section{Evaluation of community\\scoring functions}
\label{sec:ranking}
The second main purpose of the paper is to develop an evaluation methodology for network community detection. Based on ground-truth communities we now aim to compare and evaluate different community scoring functions.

\xhdr{Community goodness metrics} Our goal is to rank different structural definitions of a network community (\ie, community scoring functions) by their ability to detect ground-truth communities. We adopt the following axiomatic approach. First, we define four community ``goodness'' metrics that formalize the intuition that ``good'' communities are both compact and well connected internally while being relatively well-separated from the rest of the network.

The difference between community scoring functions from Section~\ref{sec:data} and the goodness metrics defined above is that a community scoring function quantifies how community-like a set is, while a goodness metric in an axiomatic way quantifies a desirable property of a community. A set with high goodness metric does not necessarily correspond to a community, but a set with high community score should have a high value on one or more goodness metrics. In other words, the goodness metrics shed light on various (in many cases mutually exclusive) aspects of the network community structure.

Using the notation from Section~\ref{sec:data}, we define four goodness metrics $g(S)$ for a node set $S$:
%
%
\begin{itemize}
  \denselist
  \item \textbf{Separability} captures the intuition that good communities are well-separated from the rest of the network~\cite{ShiMalik00_NCut,fortunato09community}, meaning that they have relatively few edges pointing from set $S$ to the rest of the network. Separability measures the ratio between the internal and the external number of edges of $S$: $g(S) = \frac {m_S} {c_S}$.
  \item \textbf{Density} builds on intuition that good communities are well connected~\cite{fortunato09community}. It measures the fraction of the edges (out of all possible edges) that appear between the nodes in $S$, $g(S) = \frac{m_S}{n_S(n_S-1)/2}$.
  \item \textbf{Cohesiveness} characterizes the internal structure of the community. Intuitively, a good community should be internally well and evenly connected, \ie, it should be relatively hard to split a community into two sub communities. We characterize this by the conductance of the internal cut. Formally, $g(S) = \max_{S' \subset S} \phi(S')$ where $\phi(S')$ is the conductance of $S'$ measured in the induced subgraph by $S$. Intuitively, conductance measures the ratio of the edges in $S'$ that point outside the set and the edges inside the set $S'$. A good community should have high cohesiveness (high internal conductance) as it should require deleting many edges before the community would be internally split into disconnected components~\cite{jure10community}.
  \item \textbf{Clustering coefficient} is based on the premise that network communities are manifestations of locally inhomogeneous distributions of edges, because pairs of nodes with common neighbors are more likely to be connected with each other~\cite{watts98collective}. 
\end{itemize}
%

\xhdr{Experimental setup}
We are interested in quantifying how ``good'' are the communities chosen by a particular scoring function $f(S)$ by evaluating their goodness metric. We formulate our experiments as follows: For each of 230 networks, we have a set of ground-truth communities $S_i$. For each community scoring function $f(S)$, we rank the ground-truth communities by the decreasing score $f(S_i)$. We measure the cumulative running average value of the goodness metric $g(S)$ of the top-$k$ ground-truth communities (under the ordering induced by $f(S_i)$).

The intuition for the experiments is the following. A perfect community scoring function would rank the communities in the decreasing order of the goodness metric and thus the cumulative running average of the goodness metric would decrease monotonically with $k$. While if a hypothetical community scoring function would randomly rank the communities, then the cumulative running average would be a constant function of $k$.

\xhdr{Experimental results}
We found qualitatively similar results on all our datasets. Here we only present results for the LiveJournal network. Results are representative for all other networks. We point the reader to the extended version of the paper~\cite{jaewon11comscore} for a complete set of results.

\begin{figure}[t]
	\centering
    \subfigure[Separability]{\includegraphics[width=0.235\textwidth]{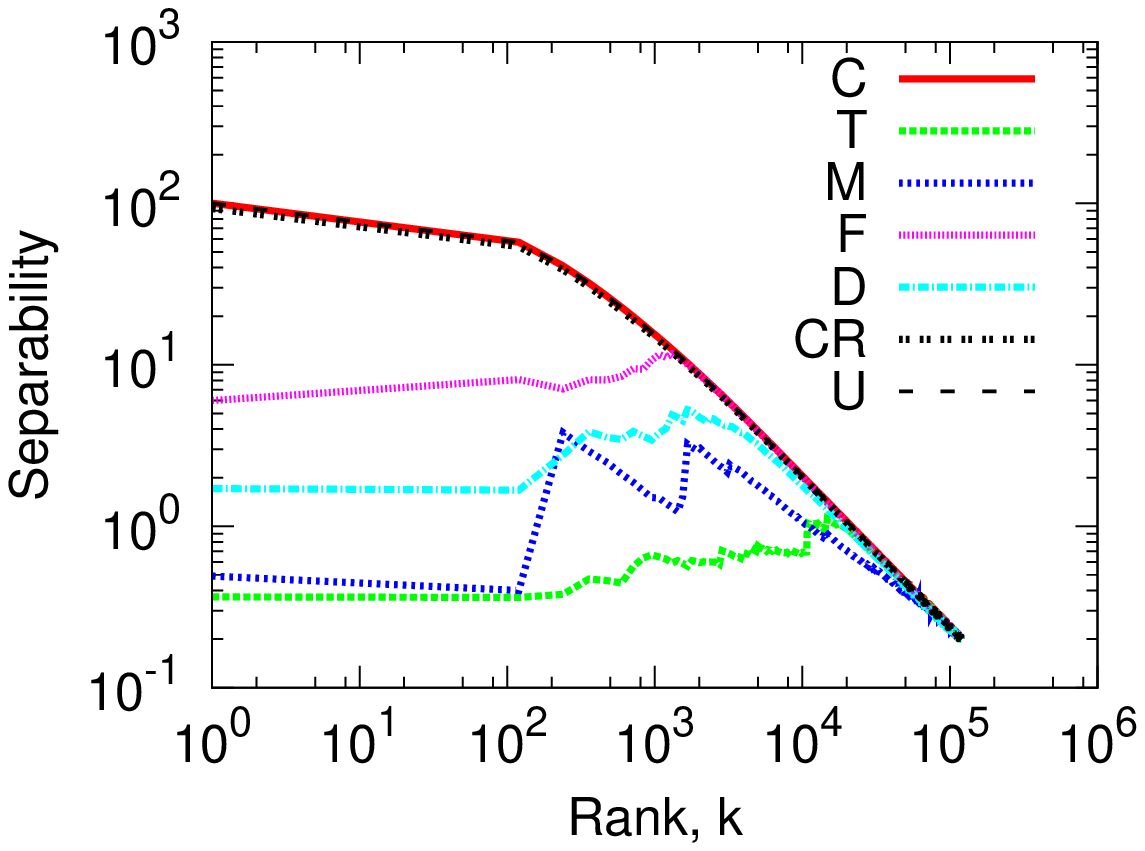}\label{fig:lj.sep}}	
    \subfigure[Density]{\includegraphics[width=0.235\textwidth]{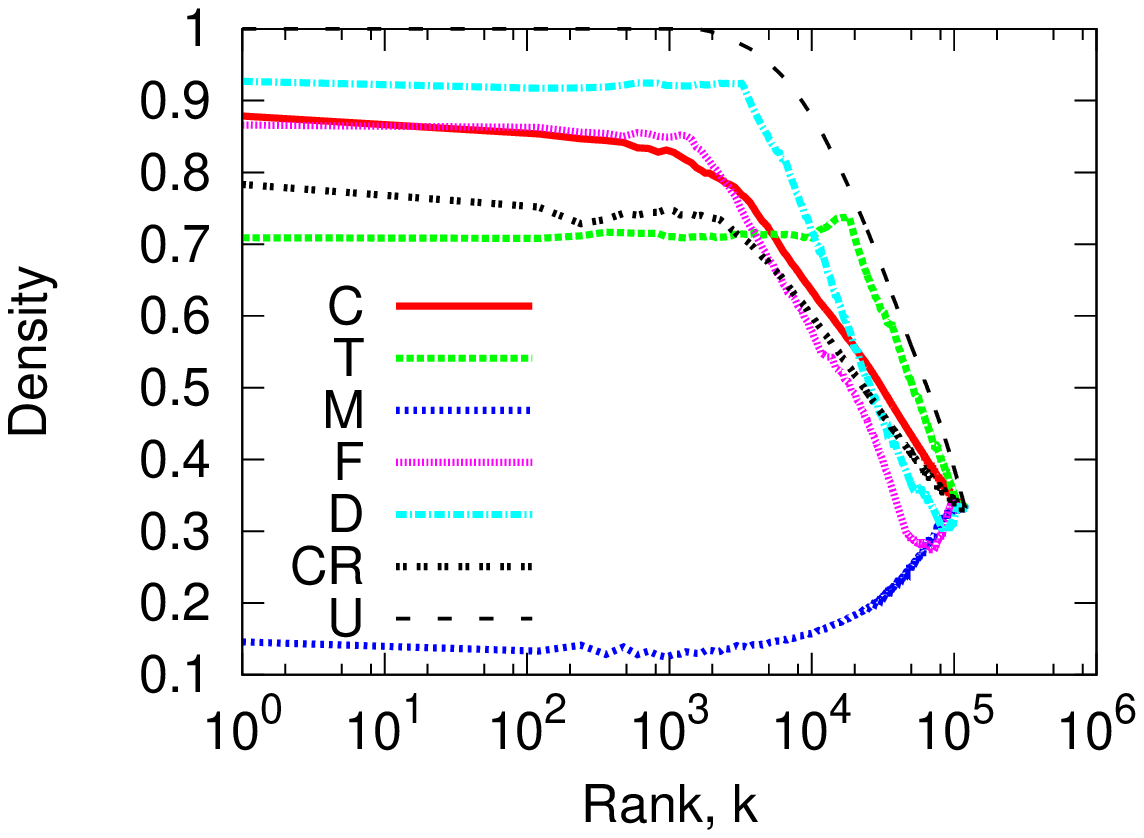}}
    \subfigure[Cohesiveness]{\includegraphics[width=0.235\textwidth]{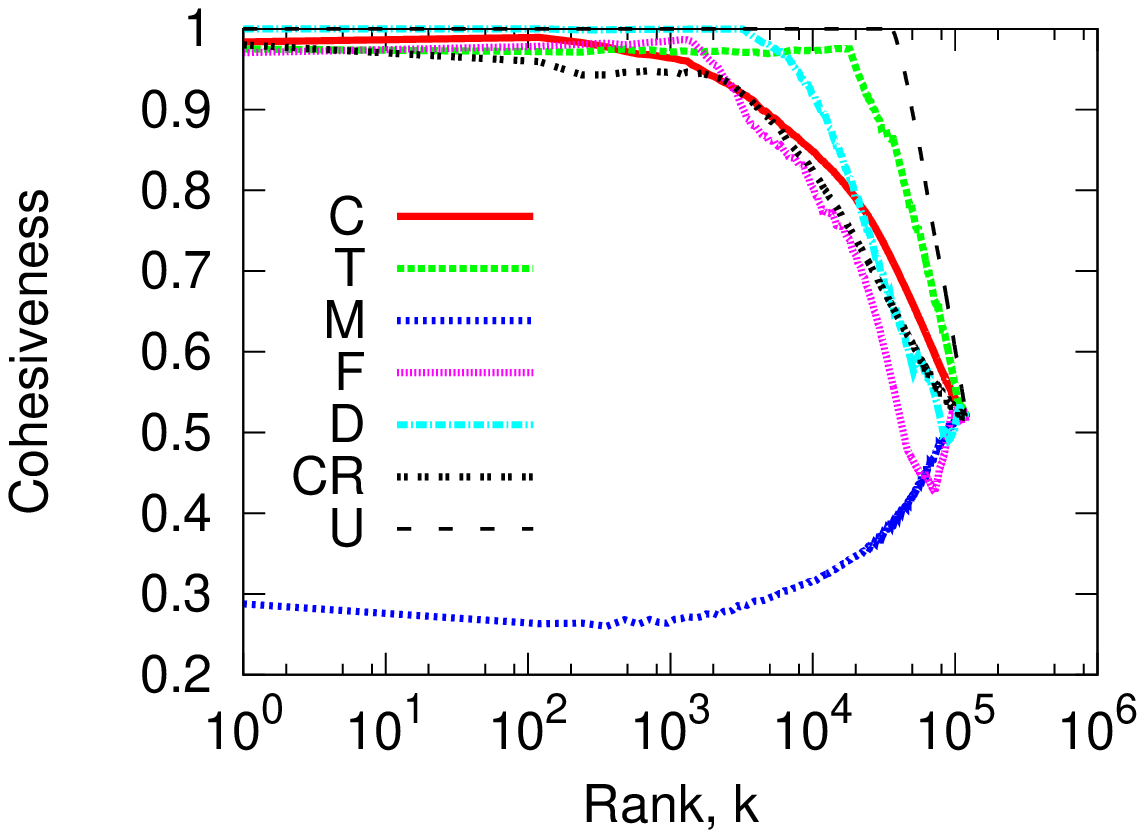}}
    \subfigure[Clustering coefficient]{\includegraphics[width=0.235\textwidth]{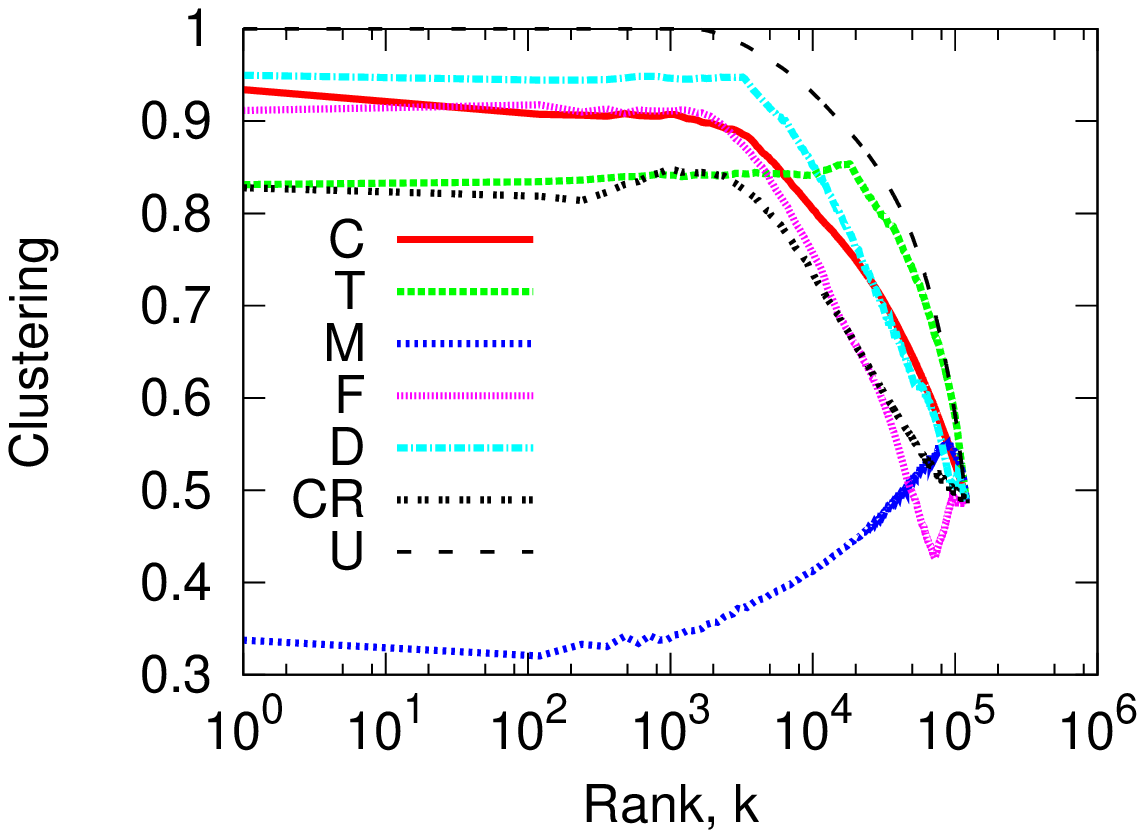}}
    \vspace{-2mm}
    \caption{Cumulative average of goodness metrics for LiveJournal communities ranked by each of the six representative scoring functions. 
    }
    \label{fig:lj.avgrank}
    \vspace{-5mm}
\end{figure}

Figure~\ref{fig:lj.sep} shows the results by plotting the cumulative running average of separability for LiveJournal ground-truth communities ranked by each of the six community scoring functions. Curve ``U'' presents the upper bound, \ie, it plots the cumulative running average of separability when ground-truth communities are ordered by decreasing separability. We observe that Conductance (C) and Cut Ratio (CR) give near optimal performance, \ie, they nearly perfectly order the ground-truth communities by separability. On the other hand, we observe that Triad Participation Ratio (T) and Modularity (M) score ground-truth communities in the inverse order of separability (especially for $k <100$), which means that they both prefer densely linked sets of nodes.



Similarly,  Figures~\ref{fig:lj.avgrank}(b), (c), and (d) show the cumulative running average of community density, cohesiveness and clustering coefficient. We observe that all scoring functions (except Modularity) rank denser, more cohesive and more clustered ground-truth communities higher. For the density metric, the Fraction over median degree (D) score performs best for high values of $k$ followed by Conductance (C) and Flake-ODF (F). In terms of cohesiveness and clustering coefficient, the Triad Participation Ratio (T) score gives by far the best results.
In all cases the only exception is the Modularity which ranks the communities in nearly reverse order of the goodness metric (the cumulative running average increases as a function of $k$). We note that these are all well-known issues of Modularity \cite{Fortunato07_ResolutionPNAS} but they get further attenuated when tested on ground-truth communities.



The curves in Figure~\ref{fig:lj.avgrank} illustrate the ability of the scoring functions to rank communities. To quantify this we perform the following experiment. For a given goodness metric $g$ and for each scoring function $f$, we measure the rank of each scoring function in comparison to other scoring functions at every value of $k$.
For example, in Figure~\ref{fig:lj.sep}, the rank at $k = 100$ of Conductance is 1, Cut ratio 2, Flake-ODF 3, FOMD 4, Modularity 5, and TPR 6. For every $k$, we rank the scores and compute the average rank over all values of $k$, which quantifies the ability of the scoring function to identify communities with high goodness metric.

\begin{table}[t]
\centering
\footnotesize
  \begin{tabular}{l||r|r|r|r}
  Scoring function &Separability&Density&Cohesiveness&Clustering\\ \hline \hline
    Conductance (C)&\textbf{1.0}&3.5&3.4&3.1\\ \hline
    Flake-ODF (F)&3.9&3.6&3.5&4.3\\ \hline
    FOMD (D)&4.9&3.0&2.9&2.9\\ \hline
    TPR (T)&4.5&\textbf{2.3}&\textbf{2.1}&\textbf{1.2}\\ \hline
    Modularity(M)&4.0&5.5&5.7&3.9\\ \hline
    CutRatio (CR)&2.6&3.1&3.2&5.5\\ \hline
  \end{tabular}
  \vspace{-2mm}
  \caption{Average scoring function rank for each goodness metric.
  }
  \vspace{-10mm}
  \label{table:score.rank}
\end{table}

Table~\ref{table:score.rank} shows the average rank for each score and each goodness metric. An average rank of 1 means that a particular score always outperforms other scores, while rank of 6 means that the score gives worst ranking out of all 6 scores. We observe that Conductance (C) performs best in terms of Separability but relatively bad in the other three metrics. For Density, Cohesiveness and Clustering coefficient, Triad Participation Ratio (T) is the best. Perhaps not surprisingly, Triad Participation Ratio scores badly on Separability of ground-truth communities. Thus, Conductance is able to identify well-separated communities, but performs poorly in identifying dense and cohesive sets of nodes with high clustering coefficient. On the other hand, Triad Participation Ratio gives the worst performance in terms of Separability but scores the best for the other three metrics.

We conclude that depending on the network different definitions of network communities might be appropriate. When the network contains well-separated non-overlapping communities, Conductance is the best scoring function. When the network contains dense heavily overlapping communities, then the Triad Participation Ratio defines the most appropriate notion of a community. Further research is needed to identify most appropriate structural definitions of communities for various types of networks and types of functional communities. E.g., in social networks we have both identity-based as well as bond-based communities~\cite{ren07bond} and they may in fact have different structural signatures.

Lastly, in Figure~\ref{fig:lj.avgrank} we also observe that the average goodness metric of the top $k$ communities remains flat but then quickly degrades. We observe the same pattern in all our data sets. Thus, for the remainder of the paper we focus our attention to a set of the top 5,000 communities of each network based on the average rank over the 6 scores.



\section{Robustness of community\\scoring functions}
\label{sec:zscores}
In this section, we evaluate community scoring functions using a set of perturbation strategies. We develop a set of strategies to generate randomized perturbations of ground-truth communities, which allows us to investigate robustness and sensitivity of community scoring functions. Intuitively, a good community scoring function should be such that it is stable under small perturbations of the ground-truth community but degrades quickly with larger perturbations.

Our reasoning is as follows. We desire a community scoring function that scores well when evaluated on a ground-truth community but scores low when evaluated on a perturbed community. In other words, an ideal community scoring function should give a maximal value when evaluated on the ground-truth community. If we consider a slightly perturbed ground-truth community (\ie, a node set that differs very slightly from the ground-truth community), we would want the score to be nearly as good as the score of the original ground-truth community. This would mean that the scoring function is robust to noise. However, if the ground-truth community is perturbed so much that it resembles a random set of nodes, then a good scoring function should give it a low score.


\xhdr{Community perturbation strategies}
We proceed by defining a set of community perturbation strategies. To vary the amount of perturbation, each perturbation strategy has a single parameter $p$ that controls the intensity of the perturbation. Given $p$ and a ground-truth community defined by its members $S$, the community perturbation starts with $S$ and then modifies it (\ie, changes its members) by executing the perturbation strategy $p|S|$ times. We define the following perturbation strategies:
\begin{itemize}
  \denselist
  \item \NodeSwap perturbation is based on the mechanism where the community memberships diffuse from the original community through the network. We achieve this by picking a random edge $(u,v)$ where $u \in S$ and $v \not\in S$ and then swap the memberships (\ie, remove $u$ from $S$ and add $v$). Note that \NodeSwap preserves the size of $S$ but if $v$ is not connected to the nodes in $S$, then \NodeSwap makes $S$ disconnected.
  \item \Random takes community members and replaces them with random non-members. We pick a random node $u \in S$ and a random $v \not\in S$ and then swap the memberships.Like \NodeSwap, \Random maintains the size of $S$ but may disconnect $S$. Generally, \Random will degrade the quality of $S$ faster than \NodeSwap, since \NodeSwap only affects the ``fringe'' of the community.
  \item \Expand perturbation grows the membership set $S$ by expanding it at the boundary. We pick a random edge $(u,v)$ where $u \in S$ and $v \not\in S$ and add $v$ to $S$. Adding $v$ to $S$ will generally decrease the quality of the community. \Expand preserves the connectedness of $S$ but increases the size of $S$.
  \item \Shrink removes members from the community boundary. We pick a random edge $(u,v)$ where $u \in S, v \not\in S$ and remove $u$ from $S$. \Shrink will decrease the quality of $S$ and result in a smaller community while preserving connectedness.
\end{itemize}
For a given $S$, let $h(S, p)$ denote a perturbed version of the community generated by the perturbation $h$ of intensity $p$.


We now quantify the difference of the score between the unperturbed ground-truth community and its perturbation. We use the Z-score, which measures in the units of standard deviation how much the scoring function changes as a function of perturbation intensity $p$. For ground-truth community $S_i$, the Z-score $Z(f, h, p)$ of community scoring function $f$ under perturbation strategy $h$ with intensity $p$ is:
\[
    Z(f,h, p) = \frac {E_i[f(S_i) - f(h(S_i, p)) ]} { \sqrt{Var_i[f(h(S_i, p))]}},
\]
where $E_i[\cdot], Var_i[\cdot]$ are the mean and the variance over communities $S_i$, and $f(h(S_i, p))$ is the community score of perturbed $S_i$ under perturbation $h$ with intensity $p$.
To measure $f(h(S_i, p))$, we run 20 trials of $h(S_i, p)$ and compute the average value of $f$. Z-score is the difference between the average community score of true communities $f(S_i)$ and the average community scores of perturbed communities $f(h(S_i, p))$ normalized by the standard deviation of community scores of perturbed communities.
%
Since $f(h(S_i, p)$ are independent for each $i$, $E_i[f(h(S_i, p)) ]$ follows a Normal distribution by the Central Limit Theorem. Thus, $P(z < Z(f, h, p))$ gives the probability that $E_i[f(h(S_i, p))] > E_i[f(S_i)]$ where $z$ is a standard normal random variable.
We measure $f$ so that lower values mean better communities, \ie, we add a negative sign to TPR, Modularity and FOMD. High Z-scores mean that $E_i[f(S_i)]$ is likely to be smaller than $E_i[f(h(S_i, p))]$ and that $S_i$ is better than $h(S_i, p)$ in terms of $f$.

\xhdr{Experimental results}
For each of the 6 community scoring functions, we measure Z-score for perturbation intensity $p$ ranging between 0.01 and 0.6. This means that we randomly swap between $1\%$ and $60\%$ of the community members and measure the Z-score for each scoring function. For small $p$, small Z-scores are desirable since they indicate that the scoring function is robust to noise. For high perturbation intensities $p$, high Z-scores are preferred because this suggests that the community scoring function is sensitive, \ie, as the community becomes more ``random'' we want the scoring function to significantly increase its value.


\begin{figure}[t]
	\centering
	\subfigure[\NodeSwap	]{\includegraphics[width=0.22\textwidth]{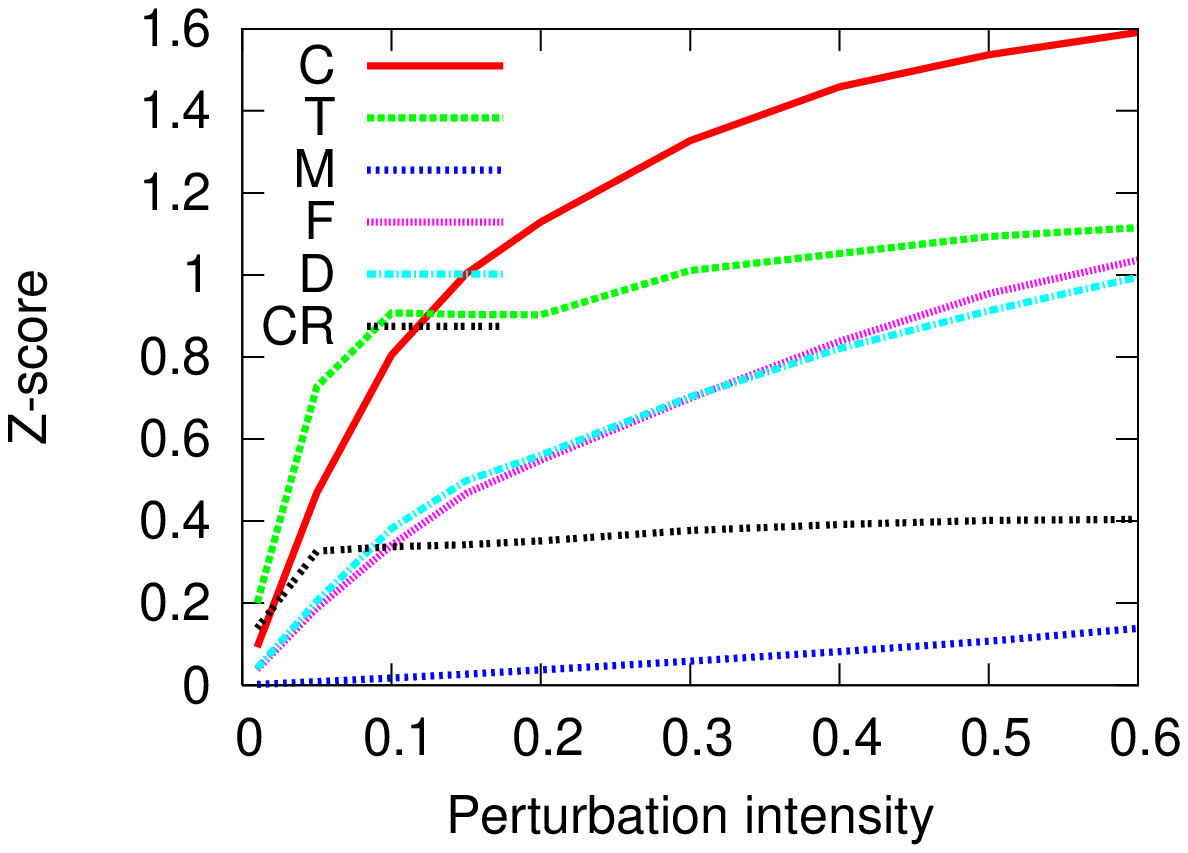}}
	\subfigure[\Random	]{\includegraphics[width=0.22\textwidth]{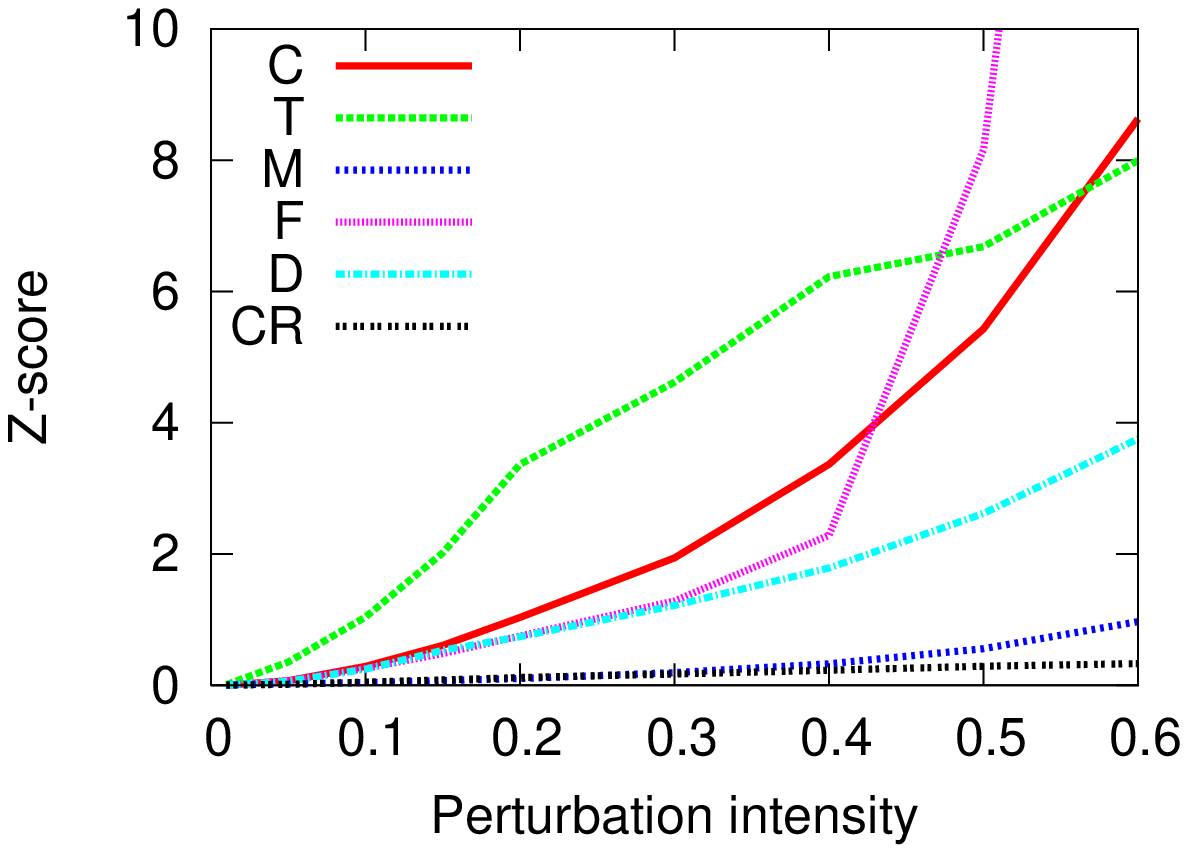}}
	\subfigure[\Expand	]{\includegraphics[width=0.22\textwidth]{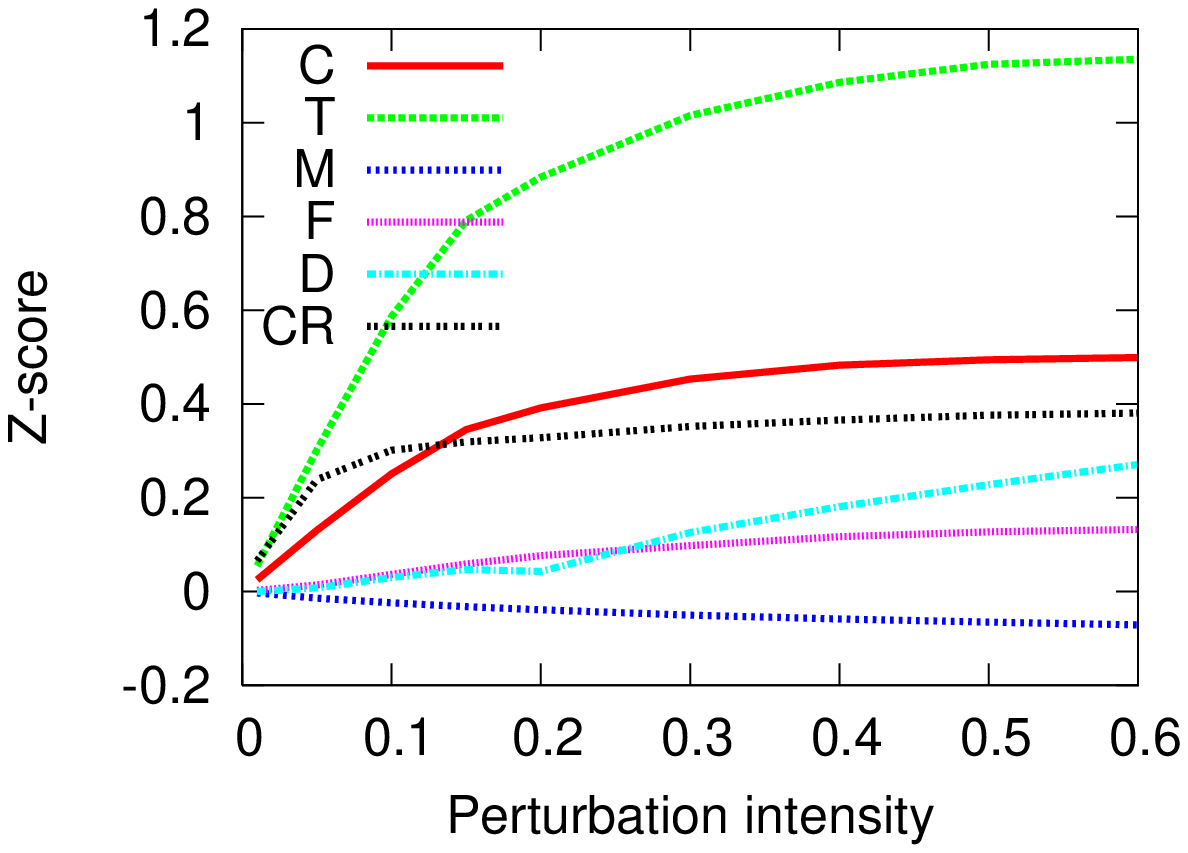} \label{fig:mod.lj.Sensitivity}}
	\subfigure[\Shrink	]{\includegraphics[width=0.22\textwidth]{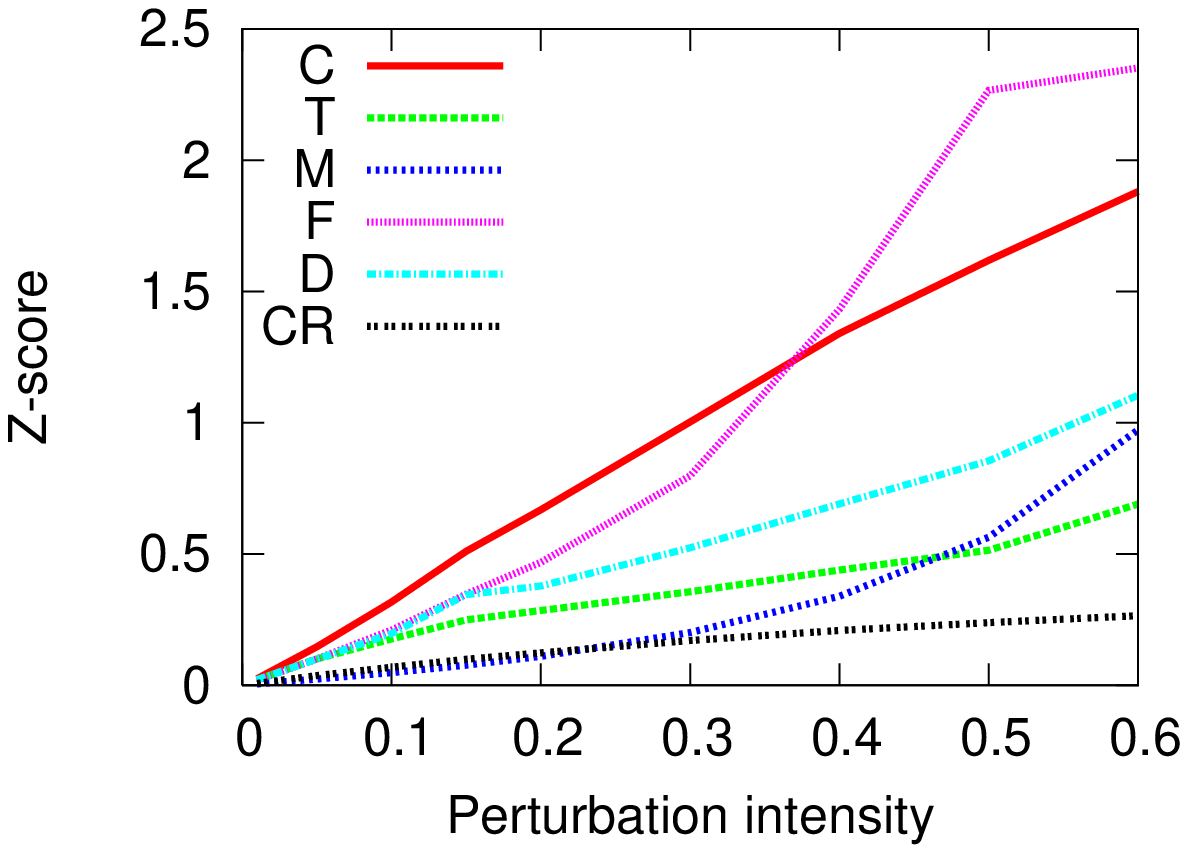}}
  \vspace{-2mm}
	\caption{Z-scores as a function of the perturbation intensity. Conductance (C) and Triad Participation Ratio (T) best detect the perturbations of LiveJournal ground-truth communities.}
\label{fig:lj.Sensitivity}
  \vspace{-3mm}
\end{figure}

Figure~\ref{fig:lj.Sensitivity} shows the Z-scores of LiveJournal communities as a function of perturbation intensity $p$. We plot the Z-score for each of the 6 community scoring functions.
As expected, the Z-scores increase with $p$, which means that as the community gets more perturbed, the value of the score tends to decrease. However, the faster the increase the more sensitive and thus the better the score.
For example, under the \NodeSwap perturbation Conductance (C) exhibits the highest Z-score after $p > 0.2$, and it has the steepest curve. Triad Participation Ratio (T) also exhibits desirable behavior. On the other hand, Modularity (M) score does not change much as we perturb the ground-truth communities. This means that Modularity is not good at distinguishing true communities from randomized sets of nodes.
%
We note very similar results on all of the remaining datasets considered in this study. Refer to the extended version for details~\cite{jaewon11comscore}.

\begin{table}[t]
\centering
\footnotesize
  \begin{tabular}{l||r|r|r|r}
    Scoring function&NodeSwap&Random&Expand&Shrink\\ \hline \hline
    Conductance (C)&\textbf{1.06}&1.59&0.50&\textbf{0.45}\\ \hline
    Flake-ODF (F)&0.51&1.15&0.11&0.41\\ \hline
    FOMD (D)&0.18&0.57&0.19&0.12\\ \hline
    TPR (T)&0.37&\textbf{1.85}&\textbf{0.74}&0.21\\ \hline
    Modularity(M)&0.23&0.14&0.03&0.15\\ \hline
    CutRatio (CR)&0.53&0.83&0.13&0.43\\
  \end{tabular}
  \vspace{-2mm}
  \caption{Average absolute increment of the Z-score between small and large community perturbations. Best performing scores are bolded.}
 \vspace{-7mm}
  \label{table:score.zslope}
\end{table}

\xhdr{Sensitivity of community scoring functions}
We also quantify the sensitivity of community scoring functions by computing the increase of the Z-score between small ($p = 0.05$) and large perturbations ($p=0.2$). As noted above, we prefer a community scoring function with fast increase of the Z-score as the community perturbation intensity increases. Table~\ref{table:score.zslope} displays the difference of the Z-score between a large and a small perturbation: $Z(f,h,0.2) - Z(f,h,0.05)$. We compute the average increment across all the 230 networks. A high value of increment means that the score is both robust and sensitive. The score is robust because even at small perturbation ($p = 0.05$) it maintains low Z-value, while at large perturbation ($p=0.2$) it has high Z-value and thus the overall Z-score increment is high.

Conductance is the most robust score under \NodeSwap and \Shrink. The Triad Participation Ratio (T) is the most robust under \Random and \Expand. In both cases  Conductance follows them closely.


\section{Discovering communities from a seed node}
\label{sec:seednodes}
Now we focus on the task of {\em inferring} communities given a single seed node. We consider two tasks that build on two different viewpoints. The first task is motivated by a community-centric view where we discover all members of community $S$ given a single member $s \in S$. The second task is motivated by a node-centric view where we want to discover {\em all} communities that a single node $s$ belongs to. This means we discover both the number of communities $s$ belongs to as well as the members of these communities.

\xhdr{Proposed method} We extend the local spectral clustering algorithm~\cite{Spielman:2004,andersen06seed} into a scalable parameter-free community detection method. The benefits of our method are: First, the method requires no input parameters and is be able to automatically detect the number of communities as well as the members of those communities. Second, the computational cost of our method is proportional to the size of the detected community ({\em not} the size of the network). Thus, our method is scalable to networks with hundreds of millions of nodes.

\begin{algorithm}[t]
  \caption{Community detection from a seed node}
  \label{alg:genLC}
  \begin{algorithmic}
  \REQUIRE Graph $G(V,E)$, seed node $s$, scoring function $f$
  \STATE {\bf (1)} Compute a random walk scores $r_u$ from seed node $s$ using PageRank-Nibble~\cite{andersen06local}.
  \STATE {\bf (2)} Order nodes $u$ by the decreasing value of $r_u/d(u)$, where $d(u)$ is the degree of $u$.
  \STATE {\bf (3)} Compute the community scoring function $f(S_k)$ of the first $k$ nodes $f_k = f( S_k = \{u_i| i \leq k\})$ for every $k$.
  \STATE {\bf (4)} Detect local minimal of $f(S_k)$ and detect one or more communities
   \IF{we want to detect one community}
    \STATE  Find the index $k^*$ at the first local optima of $f_k$.
    \RETURN $\hat{S} = \{v_i| i \leq k^*\}$
  \ELSE
      \STATE Find the indices $k^*_j$ at every local optima of $f_k$.
      \RETURN $\hat{S}_j = \{v_i| i \leq k^*_j\}$
  \ENDIF
  \end{algorithmic}
\end{algorithm}

Our method (Algorithm~\ref{alg:genLC}) builds on the findings in Sections~\ref{sec:ranking} and~\ref{sec:zscores}: First, we aim to find sets of well-connected nodes around node $s$. We achieve this by defining a local partitioning method based on random walks starting from a single seed node~\cite{andersen06local}. In particular, we use the {\em PageRank-Nibble} random walk method that computes the PageRank vector with error $< \varepsilon$ in time $O(1/\varepsilon)$ independent of the network size~\cite{andersen06seed}. The nodes with high PageRank scores from $s$ correspond to the well-connected nodes around $s$. Moreover, the random work is ``truncated'' as it sets PageRank scores $r_u$ to 0 for nodes $u$ with $r_u < \varepsilon$, for some small constant $\varepsilon$ \cite{andersen06local}. This way the computational cost is proportional to the size of the detected community and not the size of the network.



After the PageRank-Nibble assigns the proximity scores $r_u$, we sort the nodes in decreasing proximity $r_u$ and proceed to the second step of our algorithm which extends the approach of Spielman and Teng~\cite{Spielman:2004}. We evaluate the community score on a set $S_k$ of all the nodes up to $k$-th one (note that by construction $S_{k-1} \subset S_{k}$). This means that for a chosen community scoring function $f$ we compute $f(S_k)$ of the set $S_k$ that is composed of the top $k$ nodes with the highest random walk score $r_u$. The local minima of the function $f(S_k)$ then correspond to extracted communities.

We detect local minima of $f(S_k)$ using the following heuristic. For increasing $k = 1, 2, \ldots $, we measure $f(S_k)$. At some point $k^*$, $f(S_k)$ will stop decreasing and this $k^*$ becomes our ``candidate point'' for a local minimum. If $f(S_k)$ keeps increasing after $k^*$  and eventually becomes higher than $\alpha f(S_{k^*})$, we take $k^*$ as a valid local minimum. However, if $f(S_k)$ goes down again before it reaches $\alpha f(S_{k^*})$, we discard the candidate $k^*$. We experimented with several values of $\alpha$ and found that $\alpha=1.2$ gives good results across all the datasets.

\begin{figure}[t]
\centering
  \includegraphics[width=0.3\textwidth]{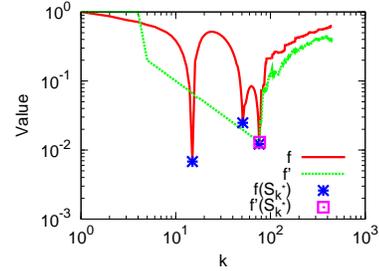}
\vspace{-2mm}
  \caption{Two community scoring functions $f$ (Conductance) and $f'$ (Triad Participation Ratio) evaluated on a set $S_k$ of top $k$ nodes with highest random walk proximity score to seed node $s$. Local optima of $f(S_k)$ correspond to detected communities. }
\vspace{-7mm}
  \label{fig:sample.ncp}
\end{figure}

For example, Fig.~\ref{fig:sample.ncp} plots $f(S_k)$ for two community scoring functions $f$ (Conductance) and $f'$ (Triad Participation Ratio). We identify the local optima (denoted by stars and squares) and use the nodes in the corresponding sets $S_k$ as the detected communities. 

Note that our method can detect multiple communities that the seed node belongs to by identifying different local minima of $f(S_k)$. However, we assume that the communities are nested (smaller communities are contained in the larger ones) even though the ground-truth communities may not necessarily follow such a nested structure. Also, note that our method is parameter-free.
 Our method differs from local graph clustering approaches~\cite{andersen06local,Spielman:2004} in two important aspects. First, instead of sweeping only using Conductance, we consider sweeping using other scoring functions. Second, we find the local optima of the sweep curve instead of the global optimum --- this change gives a large improvement over the conventional local spectral clustering approaches~\cite{andersen06local,Spielman:2004}.



\xhdr{Detecting a community from a single member}
%
%
We first consider the task where we aim to reconstruct a single ground-truth community $S$ based on one member node $s$. For each community $S$, we pick a random member node $s$ as a seed node and compare the community we detect from $s$ with the ground-truth community $S$. 
Starting from node $s$, we generate a sweep curve $f(S_k)$. Let $k^*$ be the value of $k$ where $f(S_k)$ achieves the first local minima. We then use the set $S_{k^*}$ as the detected community. Now, given the ground-truth community $S$ and the detected community $S_{k^*}$, we evaluate the precision, the recall and the F1-score. We consider 6 community scoring functions $f(\cdot)$. We compare the performance of our method to two standard community detection methods: Local Spectral clustering (LC)~\cite{andersen06local}, and the 3-clique Clique Percolation Method (CPM)~\cite{palla05_OveralpNature}.



Table~\ref{table:lc.acc} shows the performance of the proposed method for each scoring function and for the two baselines. First 5 rows show the F1-score for each of the datasets, and the last 3 rows show the average F1-score, precision and recall over all the datasets.
We observe that the Conductance (C) gives the best average F1-score, and outperforms all other scores on LiveJournal (LJ), Orkut, Amazon, and Ning. For Friendster (FS) and DBLP, the Triad participation ratio (T) performs best. This agrees with our intuition that for networks, like LiveJournal, that have fewer community overlaps scoring functions that focus on good separability perform well. In networks where nodes belong to multiple communities (like DBLP where authors publish at multiple venues), the Triad participation ratio (T) performs best. We also note that the average F1-score of Conductance is 0.46, while the baselines CPM and LC achieve F1-score of only 0.36 and 0.37, respectively. Note this is 10\% absolute and 30\% relative improvement over the state of the art baselines.


Last, we observe that some methods detect larger communities than necessary (higher recall, lower precision). Modularity (M) most severely overestimates community size. Conductance (C) and both baselines (CR and CPM) exhibit similar behavior but to a lesser extent. On the contrary, Flake-ODF (F), Fraction over median (D), Triad Participation Ratio (T), and CutRatio (CR) tend to underestimate the community size (higher precision than recall).


\begin{table}[t]
\centering
\footnotesize
  \begin{tabular}{l||c|c|c|c|c|c||c|c}
   F1-score   &C&F&D&T&M&CR&LC&CPM\\ \hline \hline
LJ&\textbf{0.64}&0.64&0.62&0.57&0.15&0.61&0.54&0.43\\ \hline
FS&0.23&0.22&0.24&\textbf{0.25}&0.24&0.18&0.13&0.14\\ \hline
Orkut&\textbf{0.21}&0.19&0.19&0.18&0.20&0.09&0.20&0.13\\ \hline
Ning &\textbf{0.24}&0.19&0.10&0.19&0.08&0.19&0.17&0.11\\ \hline
Amazon&\textbf{0.87}&0.75&0.73&0.79&0.06&0.85&0.74&0.85\\ \hline
DBLP&0.61&0.61&0.65&\textbf{0.66}&0.04&0.61&0.46&0.53\\ \hline \hline
Avg. F1 &\textbf{0.46}&0.43&0.42&0.44&0.13&0.42&0.37&0.36\\ \hline
Avg. Prec &0.50&0.53&0.52&\textbf{0.55}&0.13&0.53&0.49&0.38\\ \hline
Avg. Rec &0.60&0.47&0.51&0.47&\textbf{0.71}&0.49&0.65&0.69\\
\end{tabular}
\vspace{-2mm}
  \caption{Performance of our 6 methods and 2 baselines (LC, CPM) at detecting communities from a seed node.
}
  \vspace{-8mm}
  \label{table:lc.acc}
\end{table}



\begin{table}[t]
\footnotesize
\centering
  \begin{tabular}{c||c|c|c|c|c|c}
      $g$ &1&2&3&4&$\geq 5$&All nodes\\ \hline \hline
    LJ&0.52&0.59&0.52&0.42&0.38&0.53\\ \hline
    FS&0.13&0.10&0.08&0.05&0.02&0.13\\ \hline
    Orkut&0.21&0.17&0.13&0.11&0.10&0.20\\ \hline
    Ning (225 nets)&0.11&0.09&0.07&0.06&0.05&0.11\\ \hline
    Amazon&0.59&0.73&0.69&0.66&0.55&0.61\\ \hline
    DBLP&0.34&0.24&0.20&0.21&0.16&0.33\\
    \end{tabular}
    \vspace{-1mm}
  \caption{Average F-score between detected communities and the ground-truth communities to which a seed node belongs to, when the seed node belongs to $g$ different communities.
  }
  \label{table:multicom.acc}
  \vspace{-8mm}
\end{table}


\xhdr{Detecting all communities that a seed node belongs to}
We also explore the second task where we want to detect \emph{all} the communities to which a given seed node $s$ belongs. In this task, we are given a node $s$ that is a member of multiple communities, but we do not know which and how many communities $s$ belongs to. We detect multiple communities by detecting {\em all} the local minima (and not just the first one) of the sweep curve. This way our method both detects the number as well as the members of communities. 


For each data set, we sample a node $s$, detect communities $\hat{S}_j$, and compare them to the ground-truth communities $S_i$ that node $s$ belongs to. To measure correspondence between the true and the detected communities, we match ground-truth communities to detected communities by the Hungarian matching method~\cite{Kuhn55Hungarianbipartite}. We then compute the average F1-score over the matched pairs. We use Conductance as the community scoring function and report results in Table~\ref{table:multicom.acc}.

Note that this task is harder than the previous one as here we aim to discover multiple communities simultaneously. Whereas the previous task evaluated our method for each ground-truth community, here we first sample node $s$ and then search for the communities $S_i$ that $s$ belongs to. Therefore, larger ground-truth communities will be included in $S_i$ more often. Since larger ground-truth communities are less well separated \cite{jure08ncp2} this makes the task harder.

Table~\ref{table:multicom.acc} reports the average F1-score as a function of the number of communities $g$ that the seed node $s$ belongs to. Given that this is a harder task, we observe lower values of the F-score. Intuitively we also expect that the task becomes harder as $s$ belongs to more communities. In fact we observe that the performance degrades with increasing $g$.
 Interestingly, in LiveJournal and Amazon it appears to be easier to detect communities of nodes that belong to 2 communities than to detect a community of a node that belongs to only a single community. This is due to the fact that single community nodes reside on the border of the community and consequently Conductance produces communities that are too small~\cite{jure08ncp2}.





\section{Conclusion}
\label{sec:conclusion}
The lack of reliable ground-truth gold-standard communities has made network community detection a very challenging task. In this paper, we studied a set of 230 different large social, collaboration and information networks in which we defined the notion of ground-truth communities by nodes {\em explicitly} stating their group memberships. 

We developed an evaluation methodology for comparing network community detection algorithms based on their accuracy on real data and compared different definitions of network communities and examined their robustness. Our results demonstrate large differences in behavior of community scoring functions. Last, we also studied the problem of community detection from a single seed node. We examined class of scalable parameter-free community detection methods based on Random Walks and found that our methods reliably detect a ground-truth communities.

The availability of ground-truth communities allows for a range of interesting future directions. For example, further examining the connectivity structure of ground-truth communities could lead to novel community detection methods~\cite{jaewon12agmfit}. Overall, we believe that the present work will bring more rigor to the evaluation of network community detection, and the datasets publicly released as a part of this work will benefit the research community.

\xhdr{Acknowledgements} This research has been supported in part by NSF
IIS-1016909,		    										
CNS-1010921,		    										
CAREER IIS-1149837,											
IIS-1159679,		    										
DARPA XDATA,													
DARPA GRAPHS,													
Albert Yu \& Mary Bechmann Foundation,	
Boeing, 																
Allyes, 																
Samsung,																
Intel,                                  
Alfred P. Sloan Fellowship and 					
the Microsoft Faculty Fellowship. 			


\begin{thebibliography}{10}

\bibitem{Ahn10LinkCommunitiesNature}
Y.-Y. Ahn, J.~P. Bagrow, and S.~Lehmann.
\newblock {Link communities reveal multi-scale complexity in networks}.
\newblock {\em Nature},  Oct. 2010.

\bibitem{andersen06local}
R.~Andersen, F.~Chung, and K.~Lang.
\newblock Local graph partitioning using {PageRank} vectors.
\newblock In {\em FOCS '06}, pages 475--486, 2006.

\bibitem{andersen06seed}
R.~Andersen and K.~Lang.
\newblock Communities from seed sets.
\newblock In {\em WWW '06}, pages 223--232, 2006.

\bibitem{lars06groups}
L.~Backstrom, D.~Huttenlocher, J.~Kleinberg, and X.~Lan.
\newblock Group formation in large social networks: membership, growth, and
  evolution.
\newblock In {\em KDD '06}, pages 44--54, 2006.

\bibitem{danon05community}
L.~Danon, J.~Duch, A.~Diaz-Guilera, and A.~Arenas.
\newblock Comparing community structure identification.
\newblock {\em J. of Stat. Mech.}, 2005.

\bibitem{dhillon07graclus}
I.~Dhillon, Y.~Guan, and B.~Kulis.
\newblock Weighted graph cuts without eigenvectors: A multilevel approach.
\newblock {\em IEEE PAMI},
  29(11):1944--1957, 2007.

\bibitem{feld86focused}
S.~L. Feld.
\newblock The focused organization of social ties.
\newblock {\em Am. J. of Sociology}, 86(5):1015--1035, 1981.

\bibitem{flake00_efficient}
G.~Flake, S.~Lawrence, and C.~Giles.
\newblock Efficient identification of web communities.
\newblock In {\em KDD '00}, pages 150--160, 2000.

\bibitem{fortunato09community}
S.~Fortunato.
\newblock Community detection in graphs.
\newblock {\em Physics Reports}, 486(3-5):75 -- 174, 2010.

\bibitem{Fortunato07_ResolutionPNAS}
S.~Fortunato and M.~Barth\'{e}lemy.
\newblock Resolution limit in community detection.
\newblock {\em PNAS}, 104(1):36--41, 2007.

\bibitem{newman02community}
M.~Girvan and M.~Newman.
\newblock Community structure in social and biological networks.
\newblock {\em PNAS}, 99(12):7821--7826, 2002.

\bibitem{granovetter73ties}
M.~S. Granovetter.
\newblock The strength of weak ties.
\newblock {\em Am. J. of Sociology}, 78:1360--1380, 1973.

\bibitem{kairam12ning}
S.~Kairam, D.~Wang, and J.~Leskovec.
\newblock The life and death of online groups: Predicting group growth and
  longevity.
\newblock In {\em WSDM '12}, 2012.

\bibitem{karypis98_metis}
G.~Karypis and V.~Kumar.
\newblock A fast and high quality multilevel scheme for partitioning irregular
  graphs.
\newblock {\em SIAM Journal on Scientific Computing}, 20:359--392, 1998.

\bibitem{Kuhn55Hungarianbipartite}
H.~W. Kuhn.
\newblock The {H}ungarian method for the assignment problem.
\newblock {\em Naval Research Logistic Quarterly}, 2:83--97, 1955.

\bibitem{jure07viral}
J.~Leskovec, L.~Adamic, and B.~Huberman.
\newblock The dynamics of viral marketing.
\newblock {\em ACM TWeb}, 1(1), 2007.

\bibitem{jure10community}
J.~Leskovec, K.~Lang, and M.~Mahoney.
\newblock Empirical comparison of algorithms for network community detection.
\newblock In {\em WWW '10}, 2010.

\bibitem{jure08ncp2}
J.~Leskovec, K.~J. Lang, A.~Dasgupta, and M.~W. Mahoney.
\newblock Community structure in large networks: Natural cluster sizes and the
  absence of large well-defined clusters.
\newblock {\em Internet Mathematics}, 6(1):29--123, 2009.

\bibitem{mcpherson83blau}
M.~McPherson.
\newblock An ecology of affiliation.
\newblock {\em American Sociological Review}, 48(4):519--532, 1983.

\bibitem{mislove07measurement}
A.~Mislove, M.~Marcon, K.~P. Gummadi, P.~Druschel, and B.~Bhattacharjee.
\newblock Measurement and analysis of online social networks.
\newblock In {\em IMC '07}, pages 29--42, 2007.

\bibitem{newman2006_ModularityPNAS}
M.~Newman.
\newblock Modularity and community structure in networks.
\newblock {\em PNAS}, 103(23):8577--8582, 2006.

\bibitem{newman04community}
M.~Newman and M.~Girvan.
\newblock Finding and evaluating community structure in networks.
\newblock {\em Phys. Rev. E}, 69:026113, 2004.

\bibitem{palla05_OveralpNature}
G.~Palla, I.~Der\'{e}nyi, I.~Farkas, and T.~Vicsek.
\newblock Uncovering the overlapping community structure of complex networks in
  nature and society.
\newblock {\em Nature}, 435(7043):814--818, 2005.

\bibitem{RCCLP04_PNAS}
F.~Radicchi, C.~Castellano, F.~Cecconi, V.~Loreto, and D.~Parisi.
\newblock Defining and identifying communities in networks.
\newblock {\em PNAS}, 101(9):2658--2663, 2004.

\bibitem{ren07bond}
Y.~Ren, R.~Kraut, and S.~Kiesler.
\newblock Applying common identity and bond theory to design of online
  communities.
\newblock {\em Organization Studies}, 28(3):377--408, 2007.

\bibitem{Schaeffer07_survey}
S.~Schaeffer.
\newblock Graph clustering.
\newblock {\em Computer Science Review}, 1(1):27--64, 2007.

\bibitem{ShiMalik00_NCut}
J.~Shi and J.~Malik.
\newblock Normalized cuts and image segmentation.
\newblock {\em IEEE PAMI},
  22(8):888--905, 2000.

\bibitem{Spielman:2004}
D.~Spielman and S.-H. Teng.
\newblock Nearly-linear time algorithms for graph partitioning, graph
  sparsification, and solving linear systems.
\newblock In {\em STOC '04}, pages 81--90, 2004.

\bibitem{watts98collective}
D.~Watts and S.~Strogatz.
\newblock Collective dynamics of small-world networks.
\newblock {\em Nature}, 393:440--442, 1998.

\bibitem{jaewon12agmfit}
J.~Yang and J.~Leskovec.
\newblock Community-Affiliation Graph Model for Overlapping Network Community Detection
\newblock In {\em ICDM '12}, 2012.

\bibitem{jaewon11comscore}
J.~Yang and J.~Leskovec.
\newblock Defining and Evaluating Network Communities based on Ground-truth.
\newblock Extended version, 2012.

\end{thebibliography}
\end{document}